\documentclass[12pt]{article}
\usepackage{epsfig, url, psfrag, graphicx,color}
\usepackage[square,comma,numbers]{natbib}
\usepackage{array,amsfonts,amssymb,amsmath,longtable}
\def\sanc{\tt SANC}
\def\MTW{M_{\rm T}^{\rm W}}
\def\pTl{p_{\rm T}^{\ell}}
\begin{document}

\setcounter{page}{0}
\thispagestyle{empty}

$\,$
\vspace*{-1cm}

\begin{flushright}
{\tt IFJPAN-IV-2008-2\\
   CERN-PH-TH/2008-169\\
      June 2008      
}
\end{flushright}
\vspace*{\fill}
\begin{center}

{\LARGE\bf Implementation of SANC EW corrections \\ \vspace{2mm}
           in WINHAC Monte Carlo generator$^{\star}$} \\[2mm]
\vspace*{1.5cm}
{\bf D.~Bardin$^{ab}$, S.~Bondarenko$^{c}$, S.~Jadach$^{ad}$, 
L.~Kalinovskaya$^{ab}$, W.~P{\l}aczek$^{ea}$}

\vspace*{5mm}

{\normalsize{\it 
$^{a}$Institute of Nuclear Physics, Polish Academy of Sciences,\\
  ul.\ Radzikowskiego 152, 31-342 Cracow, Poland.\\ \vspace{1mm}
$^{b}$Dzhelepov Laboratory for Nuclear Problems, JINR,\\
       ul.\ Joliot-Curie 6, RU-141980 Dubna, Russia.    \\ \vspace{1mm}
$^{c}$Bogoliubov Laboratory of  Theoretical Physics, JINR,  \\ 
        ul.\ Joliot-Curie 6, RU-141980 Dubna, Russia.          \\ \vspace{1mm}
$^d$CERN, PH Department, CH-1211 Geneva 23, Switzerland. \\ \vspace{1mm}
$^{e}$Marian Smoluchowski Institute of Physics, Jagiellonian University,\\
   ul.\ Reymonta 4, 30-059 Cracow, Poland.
}}

\vspace*{10mm}

\end{center}

\begin{abstract}
\noindent
In this paper we describe a check of the implementation of {\tt SANC} system generated modules
into the framework of the {\tt WINHAC} Monte Carlo event generator. At this stage of work we limit
ourselves to inclusion of complete one-loop electroweak corrections 
to the charged-current Drell--Yan process.
We perform tuned comparisons of the results derived with the aid of two codes:
1) the standard {\tt SANC} integrator with YFS-inspired 
treatment of the ISR QED corrections and 
2) the {\tt WINHAC} generator, upgraded with the {\tt SANC} electroweak modules and downgraded to  
the $\cal O(\alpha)$ QED corrections.
The aim of these comparisons is to prove the correctness of implementation of the {\tt SANC}
electroweak modules into the {\tt WINHAC} generator. This is achieved through the presented tuned comparisons.

\end{abstract}

\centerline{\em Submitted to Acta Physica Polonica B}

\vspace*{5mm}

\vfill

\footnoterule
\noindent
{\footnotesize \noindent
$^{\star}$This work is partly supported by the EU grant MTKD-CT-2004-510126
 in partnership with the CERN Physics Department, by the Polish Ministry
 of Scientific Research and Information Technology grant No 620/E-77/6.PR
 UE/DIE 188/2005-2008,
 by the EU Marie Curie Research Training Network grant 
 under the contract No.\ MRTN-CT-2006-035505 
and by the Russian Foundation for Basic Research grant $N^{o}$~07-02-00932.}

\clearpage

\section{Introduction}
The main aim of this work is to implement in the Monte Carlo (MC) event generator {\tt WINHAC}~\cite{WINHAC:MC} 
the complete $\cal O(\alpha)$ electroweak
(EW) corrections delivered by the {\tt SANC} system in the form of the Standard {\tt SANC} FORTRAN Modules (SSFM) 
automatically generated by the system and to perform a cross check of this implementation by means of tuned comparisons 
of a few distributions with simple cuts. 
Here we limit ourselves to the charged current Drell--Yan-like single $W$ production and use the setup which is rooted in
the convention of TeV4LHC WS tuned comparisons working group, see Ref.~\cite{Gerber:2007xk}:
\begin{equation}
p p \;\longrightarrow\; W^+\, +\, X \;\longrightarrow\; 
\ell^+ \nu_\ell\, +\, X\,.
\end{equation}

For the description of {\tt WINHAC} and {\tt SANC} we refer the reader to the literature:
for {\tt WINHAC} to~\cite{Placzek:2003zg} and for {\tt SANC} to~\cite{Andonov:2004hi} and 
to~\cite{Bardin:2005dp}%
\footnote{{\tt SANC} is available from the project homepages at Dubna {\sf http://sanc.jinr.ru}
and CERN {\sf http://pcphsanc.cern.ch}}. 

For the case of the charged current (CC) and neutral current (NC) Drell--Yan 
(DY) processes an extended description of the {\tt SANC} approach can be found 
in Refs.~\cite{Arbuzov:2005dd} and~\cite{Arbuzov:2007db}, correspondingly.

For the final state QED radiative corrections {\tt WINHAC} has been compared with the Monte Carlo generator {\tt HORACE}, both for 
the parton-level processes and for proton--proton collisions at the LHC. Good agreement of the two programs for several 
observables has been found~\cite{CarloniCalame:2004qw}. The comparisons with generator {\tt PHOTOS} also show good 
agreement of the two generators for the QED final state radiation (FSR)~\cite{Golonka:2005pn}.

A similar event generator for the $Z$ boson production, called {\tt ZINHAC}, is under development now. 
Krakow group also works on constrained MC algorithms for the QCD ISR parton shower that could be applied to Drell--Yan
processes, see, e.g.\ Ref.~\cite{Jadach:2007singleCMC}.

Many results of tuned comparison of {\tt SANC} with several other programs were presented for CC case in 
Ref.~\cite{Buttar:2006zd} and~\cite{Gerber:2007xk} and for NC case in~\cite{Buttar:2008jx}, showing very 
good agreement. This ensures us in a high confidence of NLO EW {\tt SANC} predictions.

In this paper we limit ourselves to presenting the numerical tests of the 
implementation of {\tt SANC} EW corrections in generator {\tt WINHAC}, detailed description of
the implementation itself will be given elsewhere. The paper is organized 
as follows. In Section~2 we describe the setup of the tuned comparisons
between {\tt SANC} and {\tt WINHAC}. In Section~3 we present the results of these 
comparisons for the total cross sections and various distributions, first
at the Born level then for ${\cal O}(\alpha)$ EW corrections, and finally for a model of
purely weak corrections. Finally, Section~4 concludes the paper.

\section{Setup of tuned comparisons of {\tt SANC} and {\tt WINHAC}\label{sec:setup}}
\noindent
We use the input parameter set as in Ref.~\cite{Gerber:2007xk}, see also comments after Eq.~(4.4.37):
\begin{eqnarray}\label{eq:pars}
G_{\mu} = 1.16637\times 10^{-5} \; {\rm GeV}^{-2}, 
& \quad & \alpha= 1/137.03599911, \quad \alpha_s(M_Z^2)=0.1176, 
\nonumber \\ 
M_Z = 91.1876 \; {\rm GeV}, & \quad & \Gamma_Z =  2.4924  \; {\rm GeV},
\nonumber  \\
M_W = 80.37399 \; {\rm GeV}, & \quad & \Gamma_W = 2.0836 \; {\rm GeV},
\nonumber  \\
M_H = 115 \; {\rm GeV}, & \quad & 
\nonumber  \\
m_e  = 0.51099892 \; {\rm MeV}, &\quad &m_{\mu}=0.105658369 \; {\rm GeV},  
\nonumber  \\
\quad m_{\tau}=1.77699 \; {\rm GeV},
\nonumber  \\
m_u=0.06983 \; {\rm GeV}, & \quad & m_c=1.2 \; {\rm GeV}, 
\quad m_t=174 \; {\rm GeV},
\nonumber  \\
m_d=0.06984 \; {\rm GeV}, & \quad & m_s=0.15 \; {\rm GeV}, \quad m_b=4.6 \; {\rm GeV}, 
\nonumber \\
|V_{ud}| = 0.975, & \quad & |V_{us}| = 0.222, 
\nonumber \\
|V_{cd}| = 0.222, & \quad & |V_{cs}| = 0.975,
\nonumber \\
|V_{cb}|=|V_{ts}|=|V_{ub}|\hspace*{-2mm}&=&\hspace*{-2mm} |V_{td}|= |V_{tb}|=0.  
\end{eqnarray}
However, we present the results both in the $\alpha(0)$ and $G_\mu$ one-loop parametrization schemes.

To compute the hadronic cross section we also use the MRST2004QED set of parton density 
functions~\cite{Martin:2004dh}, and take the renormalization scale, $\mu_r$, and the QED and QCD factorization
scales, $\mu_{\rm QED}$ and $\mu_{\rm QCD}$, to be $\mu_r^2=\mu_{\rm QED}^2=\mu_{\rm QCD}^2=M_W^2$.

We impose only detector acceptance cuts on the leptons transverse momenta and the charged lepton pseudorapidity ($\eta_\ell$):
\begin{equation}
\pTl>20~{\rm GeV,}\qquad\qquad |\eta(\ell)|<2.5, \qquad\qquad
\ell=e,\,\mu ,
\label{eq:lepcut}
\end{equation}
\begin{equation}
p\llap/_{\rm T}^\nu>20~{\rm GeV,}
\label{eq:ptmisscut}
\end{equation}
where $p\llap/_{\rm T}^\nu$ is the missing transverse momentum originating from the neutrino. 

To simplify the conditions of this purely technical comparison, we do not impose lepton identification requirements, 
as given in Table~4.4.49 of Ref.~\cite{Gerber:2007xk}, so we provide ``simplified bare'' results, i.e.\ without
smearing, recombination and lepton separation cuts.
We present our results only for three differential distributions and the total cross sections,
at LO and NLO, and the corresponding relative corrections, $\delta_{\rm EW}\,[\%]=d\sigma_{\rm NLO}/d\sigma_{\rm LO}-1$,
for two processes: $pp \to W^+ + X \to \ell^+\nu_\ell + X$ with $\ell=e,\,\mu$ at the LHC in two schemes: $\alpha(0)$ and $G_\mu$.
Moreover, we present the results for some well-defined model of ``purely weak'' corrections $\delta_{\rm weak}$, 
given in Subsection~\ref{subsec:weak}, for the same cases as for $\delta_{\rm EW}$. 

\vspace*{5mm}

In our comparisons we use the following {\it $W$-boson observables}: 
\begin{itemize}
\item
$\sigma_W$: the {\em total inclusive cross section} of the $W$-boson production.\\
\item
$\displaystyle\frac{d\sigma}{d\MTW}$: the {\em transverse mass} distribution of the lepton lepton--neutrino pair.\\
\noindent
The transverse mass is defined as
\begin{equation}
\MTW=\sqrt{2\pTl p_{\rm T}^\nu(1-\cos\phi^{\ell\nu})}\,,
\label{eq:mt}
\end{equation}
where $p_{\rm T}^\nu$ is the transverse momentum of the neutrino, and $\phi^{\ell\nu}$ is the angle between the charged 
lepton and the neutrino in the transverse plane. The neutrino transverse momentum is
identified with the missing transverse momentum, $p\llap/_{\rm T}$, in the event.
\item
$\displaystyle\frac{d\sigma}{d\pTl}$: the {\em transverse lepton momentum} distribution. \\
\item
$\displaystyle\frac{d\sigma}{d|\eta_\ell|}$: the {\em lepton pseudorapidity} distribution \\
\begin{equation}
\eta_\ell=-\ln\left(\tan\frac{\theta_\ell}{2}\right),
\end{equation}
where the lepton kinematical variables are defined in the laboratory frame.\\
\end{itemize}

One should emphasize an important difference between the conditions of these comparisons and that of 
TeV4LHC WS concerning the subtraction of initial quark mass singularities. Instead of the commonly adopted 
$\overline{\rm MS}$ or DIS subtraction scheme (as, for example, in Ref.~\cite{Gerber:2007xk}),
we use here an YFS-inspired subtraction method~\cite{Wackeroth:1996hz}. 
\begin{equation}
d\sigma^{\rm YFS}_{\rm ISR}(\hat{s},m_d,m_u;\epsilon)=d\sigma^{\rm Born}_{\rm ISR}(\hat{s},m_d,m_u;\epsilon)\,
                                                      \delta^{\rm YFS}_{\rm ISR}(\hat{s},m_d,m_u;\epsilon),
\label{eq:sigma_yfs_subtr}
\end{equation}
where
\begin{eqnarray}
\delta^{\rm YFS}_{\rm ISR}(\hat{s},m_d,m_u;\epsilon)&=&\frac{\alpha}{\pi}\biggl\{\biggl[
 Q_d^2\biggl(\ln\frac{\hat{s}}{m^2_d}-1\biggr)
+Q_u^2\biggl(\ln\frac{\hat{s}}{m^2_u}-1\biggr)-1\biggr]\ln\epsilon
\\ \nonumber
&&+Q_d^2\biggl(\frac{3}{4}\ln\frac{\hat{s}}{m^2_d}-1+\frac{\pi^2}{6}\biggr)
  +Q_u^2\biggl(\frac{3}{4}\ln\frac{\hat{s}}{m^2_u}-1+\frac{\pi^2}{6}\biggr)
  +1-\frac{\pi^2}{3}\biggr\},
\label{eq:delta_yfs_subtr}
\end{eqnarray}
with
\begin{equation}
\epsilon = \frac{2\omega}{\sqrt{\hat{s}}}
\end{equation}
being the dimensionless soft--hard photon separator 
($\omega$ is the photon energy). 
The $Q_u,\,Q_d$ are the electric charges of the up-type and down-type quarks 
in the units of the positron charge and $m_u,\,m_d$ are their masses,
while $\hat{s}$ is the centre-of-mass energy squared of the incoming quarks.

Simultaneously, we subtract in a gauge-invariant way the contribution of 
the ISR hard photons, 
derived using the $W$ propagator splitting technique~\cite{Berends:1984xv}.
In this way the initial quark mass dependence drops out from the one-loop level observables.

In order to define our ``weak'' corrections, we will need the YFS corrections for the ``initial-final'' interference 
\begin{eqnarray}
\label{eq:delta_yfs_int}
\delta^{\rm YFS}_{\rm Int}(\hat{s},t,u;\epsilon)&=&\frac{\alpha}{\pi}\biggl\{
2\biggl[Q_d\ln\frac{\hat{s}}{-t}-Q_d\ln\frac{\hat{s}}{-u}+1\biggr]
\ln\frac{M^2_W\,\epsilon}{\sqrt{\left(s-M^2_W\right)^2+M^2_W\Gamma^2_W}}
\\ \nonumber
&&+Q_d\biggl[\frac{1}{2}\ln\frac{\hat{s}}{-t}\biggl(\ln\frac{\hat{s}}{-t}+1\biggr)+{\rm Li}_2\left(1+\frac{\hat{s}}{t}\right)\biggr]
\\ \nonumber
&&-Q_u\biggl[\frac{1}{2}\ln\frac{\hat{s}}{-u}\biggl(\ln\frac{\hat{s}}{-u}+1\biggr)+{\rm Li}_2\left(1+\frac{\hat{s}}{u}\right)\biggr]
+\frac{\pi^2}{6}-2\biggr\},
\end{eqnarray}
and for the ``final state radiation''
\begin{eqnarray}
\label{eq:delta_yfs_fsr}
\delta^{\rm YFS}_{\rm FSR}(\hat{s},m_l;\epsilon)&=&\frac{\alpha}{\pi}\biggl\{
 \biggl(\ln\frac{\hat{s}}{m^2_l}-2\biggr)\ln\epsilon+\frac{3}{4}\ln\frac{\hat{s}}{m^2_l}-\frac{\pi^2}{6}\biggr\},
\end{eqnarray}
where $\hat{s},\,t,\,u$ are the standard Mandelstam variables for the
parton-level process 
and $m_l$ is the charged lepton mass.

\begin{figure}[!ht]
\begin{center} 
\includegraphics[width=12.5cm]{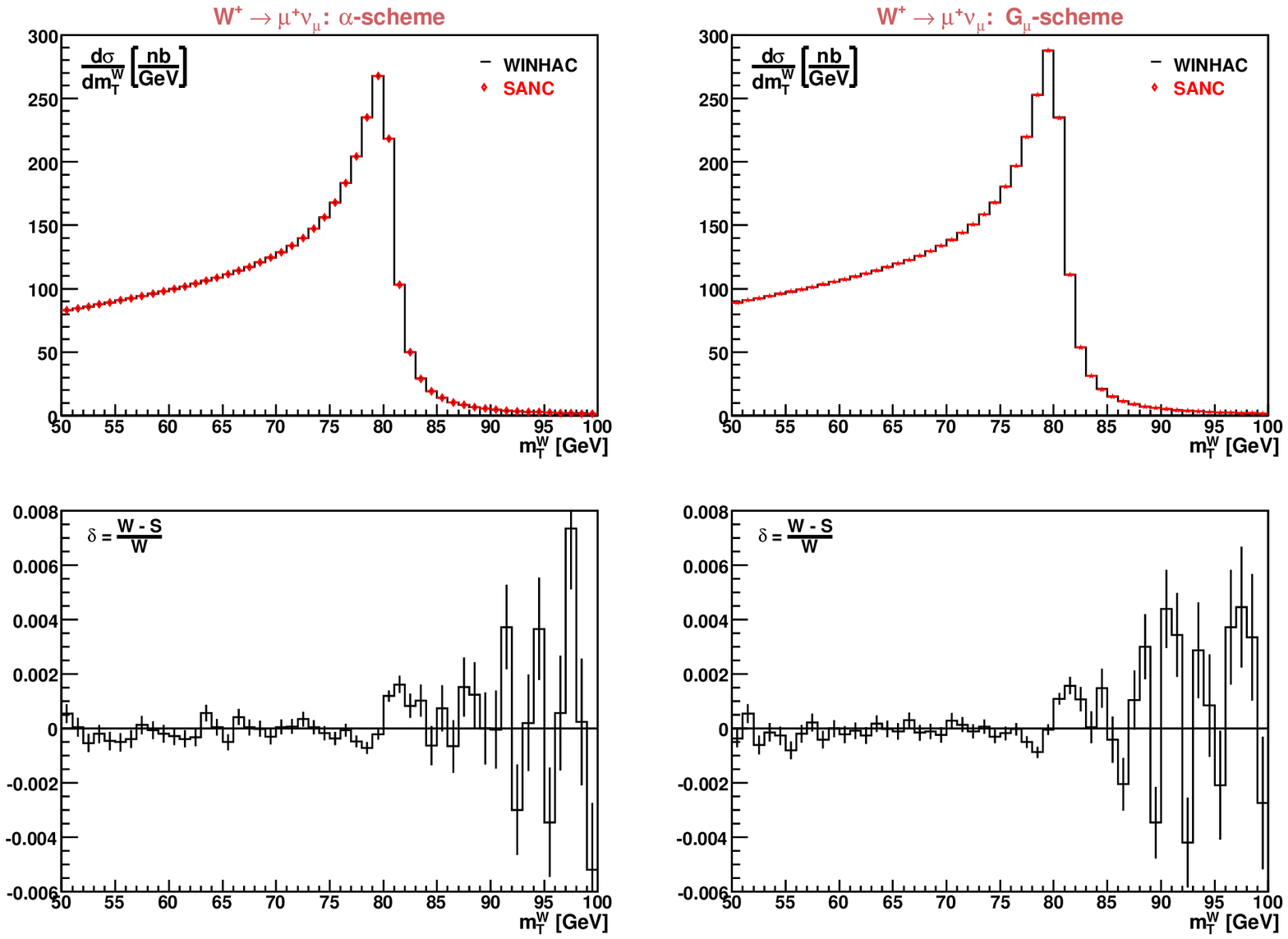}
\end{center}
\vspace*{-5mm}
\caption{The Born distributions of $\MTW$  from {\tt SANC} (red diamonds) and {\tt WINHAC} (solid lines) in two schemes and their relative deviations $\delta=\frac{{\rm W}-{\rm S}}{{\rm W}}.$
}\label{fig:mtmutree}
\end{figure}

\section{Numerical results}\label{sec:comp}

In this section we present the numerical results of the tuned comparisons 
between {\tt SANC} and {\tt WINHAC}, first the the Born level (LO) and then
including the ${\cal O}(\alpha)$ EW corrections (NLO). At the end of this
section we compare also the so-called ``purely weak'' corrections which are 
the difference between the EW corrections and the ``QED'' corrections 
defined by the terms given in 
Eqs.~(\ref{eq:delta_yfs_int}--\ref{eq:delta_yfs_fsr}) 
plus the corresponding hard-photon contributions.

\subsection{Comparisons at tree level, LO}
\begin{figure}[!ht]
\begin{center} 
\includegraphics[width=12.5cm]{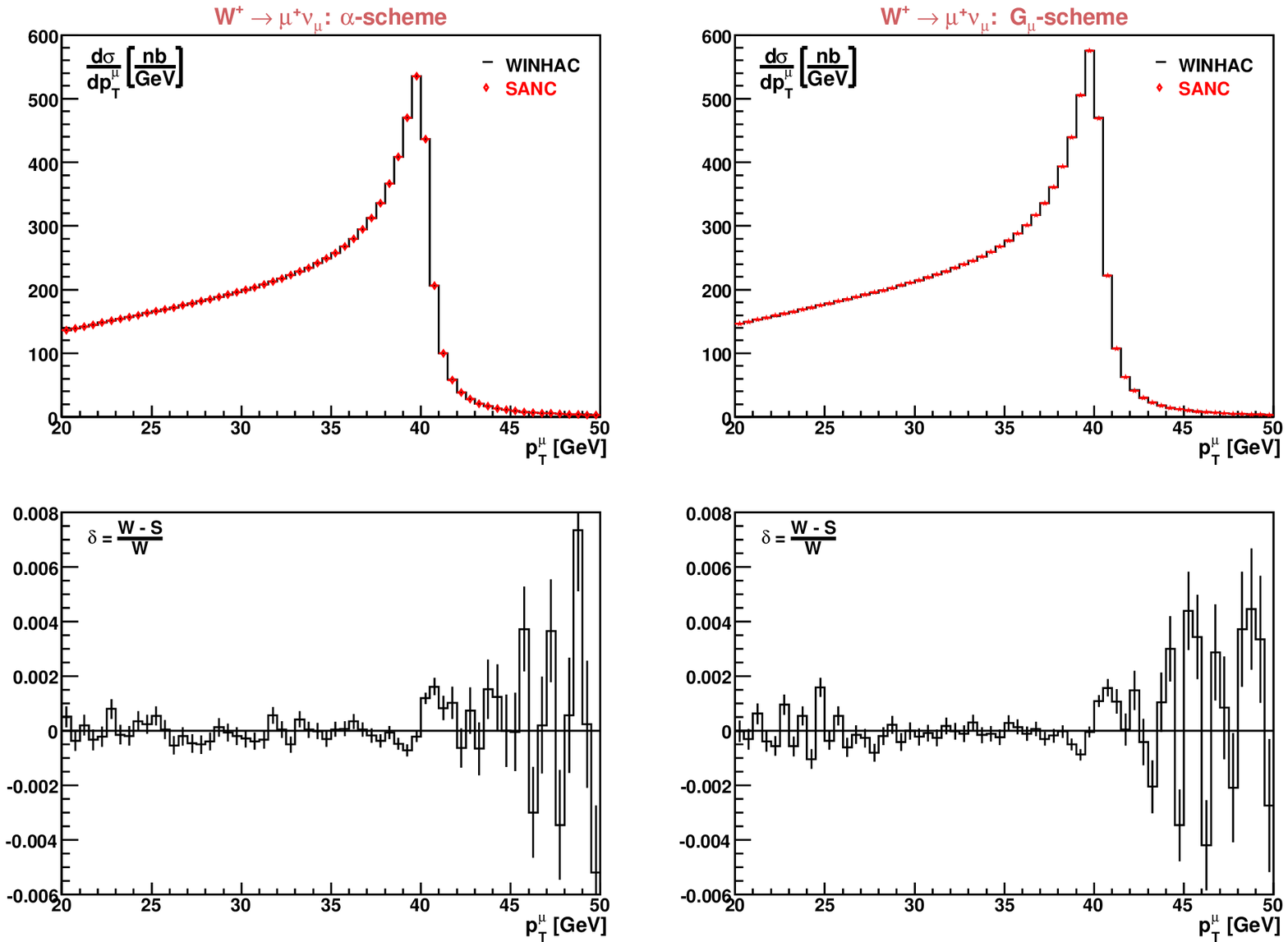}
\end{center}
\vspace*{-5mm}
\caption{The Born distributions of $\pTl$ from {\tt SANC} (red diamonds) and {\tt WINHAC} (solid lines) in two schemes and their relative deviations $\delta=\frac{{\rm W}-{\rm S}}{{\rm W}}.$
}\label{fig:ptmutree}
\end{figure}

\begin{figure}[!ht]
\begin{center}
\includegraphics[width=12.5cm]{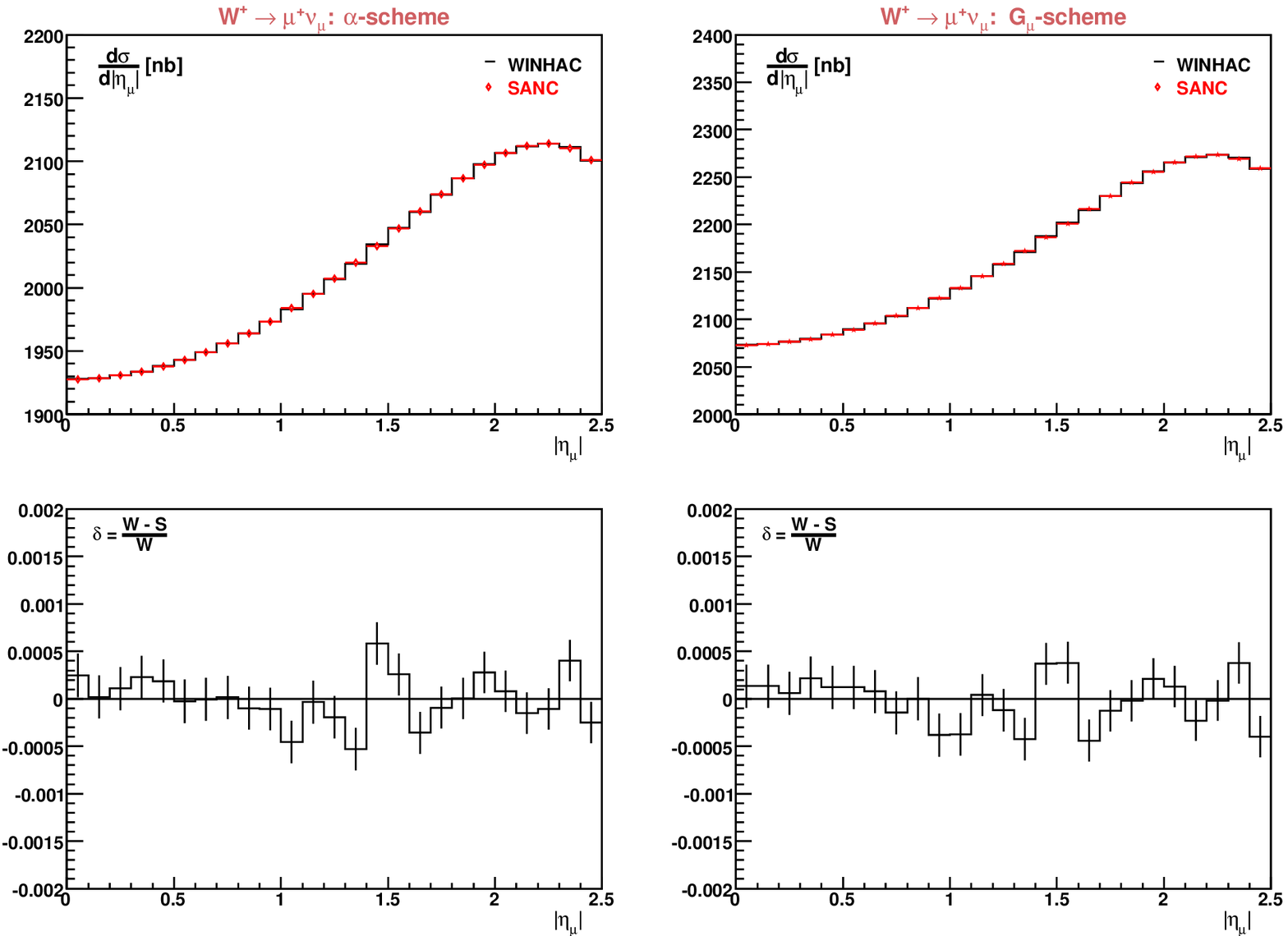}
\end{center}
\vspace*{-5mm}
\caption{The Born distributions of $|\eta_\ell|$ from {\tt SANC} (red diamonds) and {\tt WINHAC} (solid lines) in two schemes and their relative deviations $\delta=\frac{{\rm W}-{\rm S}}{{\rm W}}.$
}\label{fig:etamutree}
\end{figure}

We begin with the comparisons at the Born level. In Figs.~\ref{fig:mtmutree}--\ref{fig:etamutree}
the distributions are shown for all three observables under consideration only for $\mu^{+}$ final state
but in the both schemes: $\alpha(0)$ and $G_{\mu}$. The lower parts of the figures shows the relative deviation 
$\Delta=({\rm W}-{\rm S})/{\rm W}$
between the two calculations (${\rm W}$ for {\tt WINHAC}, ${\rm S}$ for {\tt SANC}).


As seen, the relative deviations lie within the $1$ per-mill band, 
wherever the cross section is not very small%
\footnote{On the {\tt SANC} side we have both a VEGAS~\cite{Lepage:1977sw} based integrator and a
FOAM~\cite{Jadach:2005ex} based event generator. In this comparison the integrator has been used.}. 
We do not show the comparisons for electron channel, since at tree level the muon mass effects are negligible, and the plots look identical.

\subsection{Comparison at one-loop level, NLO inclusive cross sections}

Turning to the NLO results, we show, first of all, in Table~\ref{tab:comparison} 
the comparisons of the inclusive cross sections 
(in pb) within the acceptance cuts and the relative radiative correction factor (in \%), as seen by two calculations 
(second and third rows). In the first row we show {\tt SANC} results in the conditions of TeV4LHC WS.
The numbers agree with those published in \cite{Gerber:2007xk} within statistical errors.

\begin{table}[!ht]
\begin{center}
\begin{tabular}{||l|l|l|l|l|l|l||}
\hline
\hline
\multicolumn{7}{||c||}{\bf LHC, $p p \to W^+\,+\,X \to e^+ \nu_e\,+\,X$}                       \\ \hline
& \multicolumn{3}{|c|}{$\alpha$-scheme} & \multicolumn{3}{|c||}{$G_{\mu}$-scheme} \\ \hline
& LO [pb] & NLO [pb] & $\delta_{\rm EW}$ [\%] & LO [pb] & NLO [pb] & $\delta_{\rm EW}$ [\%] \\ 
\hline
{\tt SANC-$\overline{\tt MS}$}&$5039.19(2)$ &$5139.33(5) $& $1.987(1)$ &     ---     &     ---     &    ---     \\ 
{\tt SANC-YFS}                &$5039.19(2)$ &$5137.53(3) $& $1.952(1)$ &$5419.18(2)$ &$5208.48(3) $&$-3.888(1)$ \\
{\tt WINHAC}                  &$5039.06(11)$&$5138.04(16)$& $1.966(3)$ &$5419.04(12)$&$5209.04(12)$&$-3.874(3)$  \\
\hline\hline
\multicolumn{7}{||c||}{\bf LHC, $p p \to W^+\,+\,X \to \mu^+ \nu_\mu\,+\,X$}                   \\ \hline
& \multicolumn{3}{|c|}{$\alpha$-scheme} & \multicolumn{3}{|c||}{$G_{\mu}$-scheme} \\ \hline
& LO [pb] & NLO [pb] & $\delta_{\rm EW}$ [\%] & LO [pb] & NLO [pb] & $\delta_{\rm EW}$ [\%] \\ 
\hline
{\tt SANC-$\overline{\tt MS}$}&$5039.20(2)$ &$5229.58(6)$ & $3.778(1)$ &     ---     &     ---     &    ---     \\ 
{\tt SANC-YFS}                &$5039.20(2)$ &$5227.73(2)$ & $3.741(1)$ &$5419.19(2)$ &$5305.47(3)$ &$-2.098(1)$ \\ 
{\tt WINHAC}                  &$5039.03(11)$&$5227.87(14)$& $3.745(2)$ &$5419.01(12)$&$5305.59(14)$&$-2.094(2)$ \\
\hline\hline
\end{tabular}
\caption{The tuned comparisons of the LO and EW NLO predictions for $\sigma_{W}$ and {$\delta_{\rm EW}$}
from {\tt SANC} and {\tt WINHAC} for the simplified bare cuts.
The statistical errors of the Monte Carlo integration are given in parentheses.} 
\label{tab:comparison}
\end{center}
\end{table}

\begin{figure}[!ht]
\begin{center} 
\includegraphics[width=12.0cm]{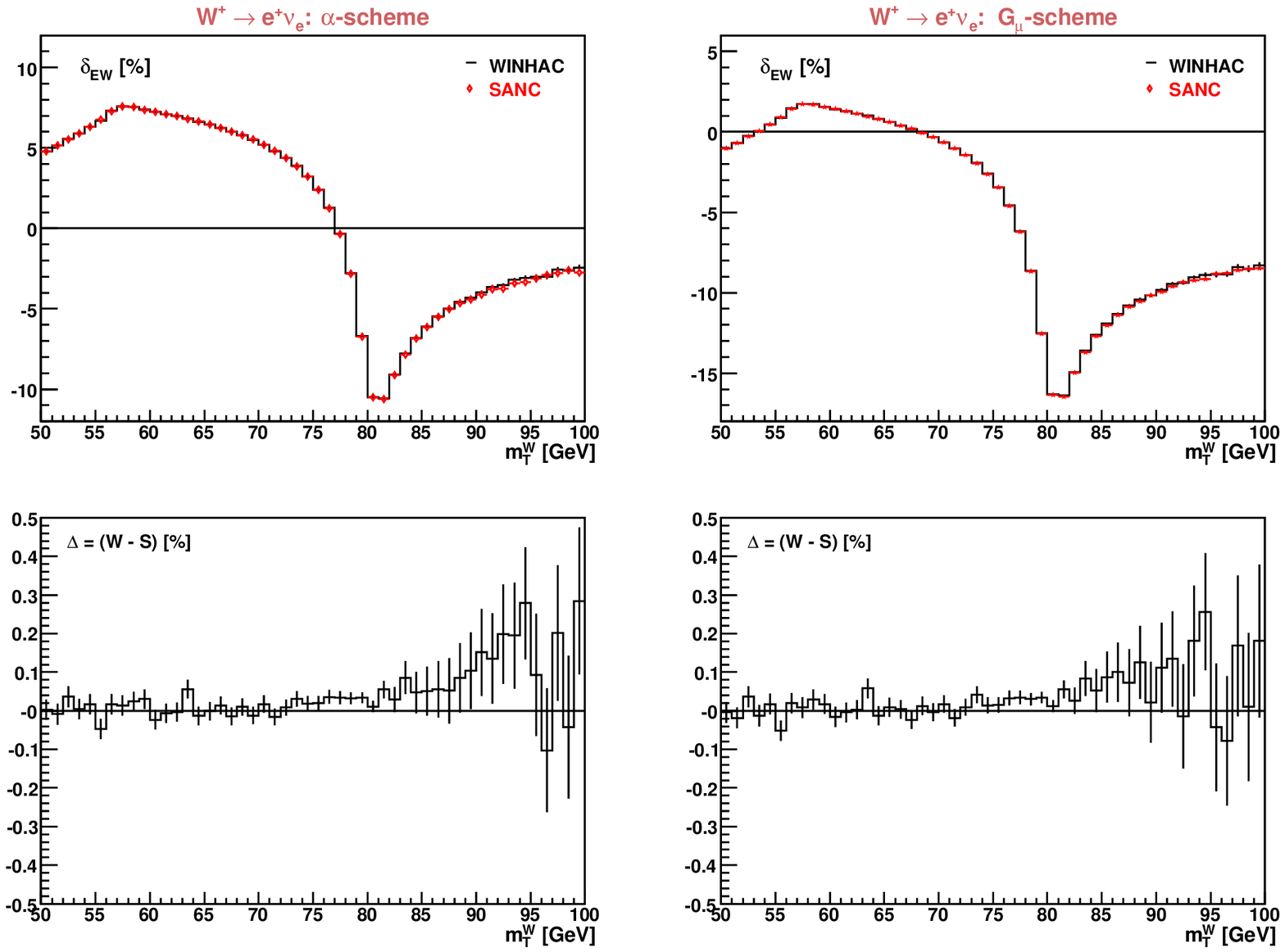}
\end{center}
\vspace*{-5mm}
\caption{The EW NLO distributions of $\MTW$ from {\tt SANC} (red diamonds) and {\tt WINHAC} (solid lines) for the electron channel in two schemes and their absolute deviations 
$\Delta={{\rm W}-{\rm S}}$.
}\label{fig:mtel1loop}
\end{figure}

The Born cross sections  from {\tt SANC} and {\tt WINHAC} agree well within statistical errors ($<10^{-4}$). 
The EW NLO cross sections agree not worse than within a half a per mill or agree even within statistical errors in both 
schemes, both for the electron and muon channels, better for the muon channel where we observe the agreement within 
the statistical errors.

\subsubsection{NLO distributions: electron channel}
\begin{figure}[!ht]
\begin{center} 
\includegraphics[width=12.0cm]{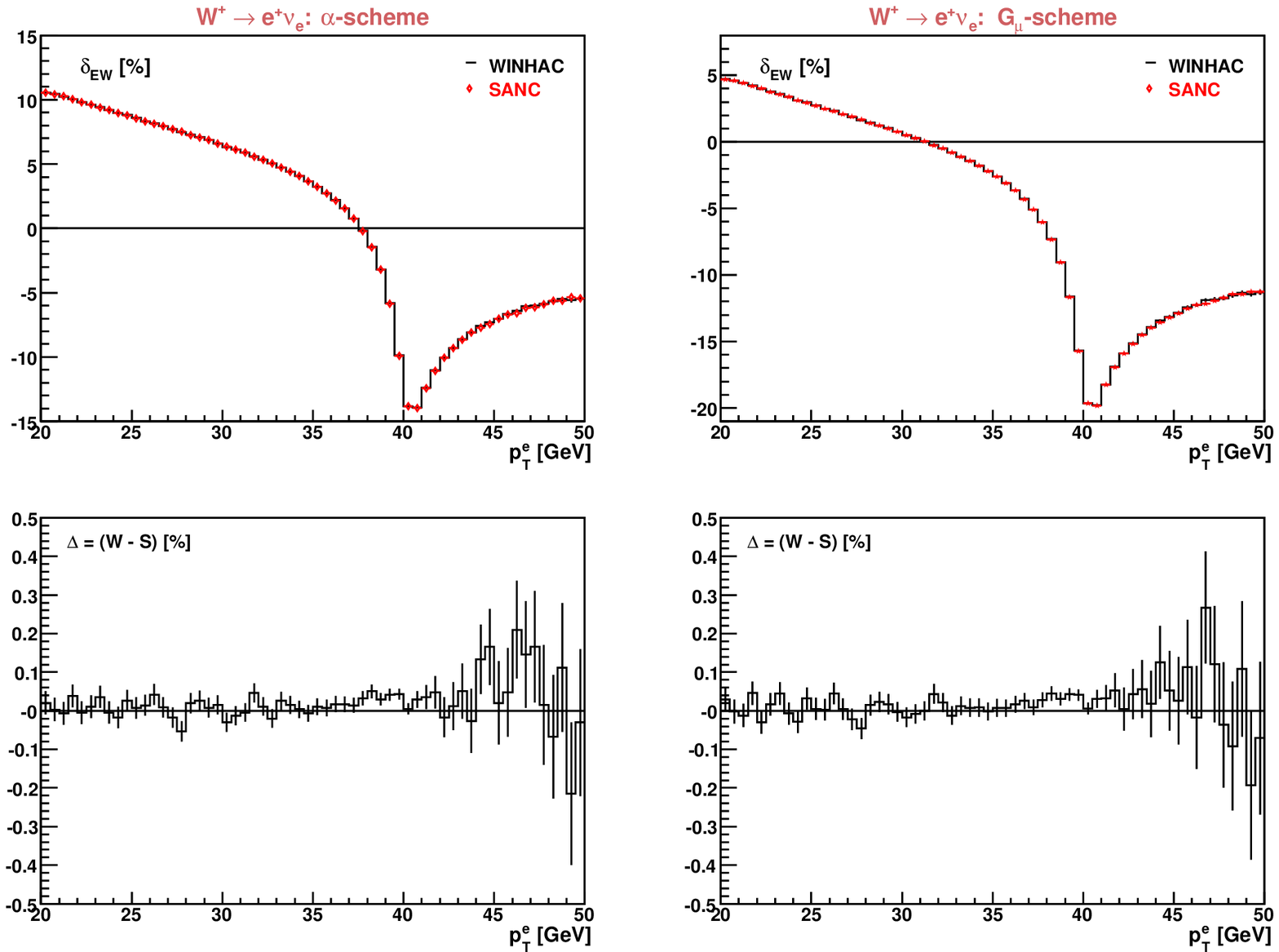}
\end{center}
\vspace*{-5mm}
\caption{The EW NLO distributions of $\pTl$ from {\tt SANC} (red diamonds) and {\tt WINHAC} (solid lines) for the electron channel in two schemes and their absolute deviations $\Delta={{\rm W}-{\rm S}}$.
}\label{fig:ptel1loop}
\end{figure}

\begin{figure}[!ht]
\begin{center} 
\includegraphics[width=12.0cm]{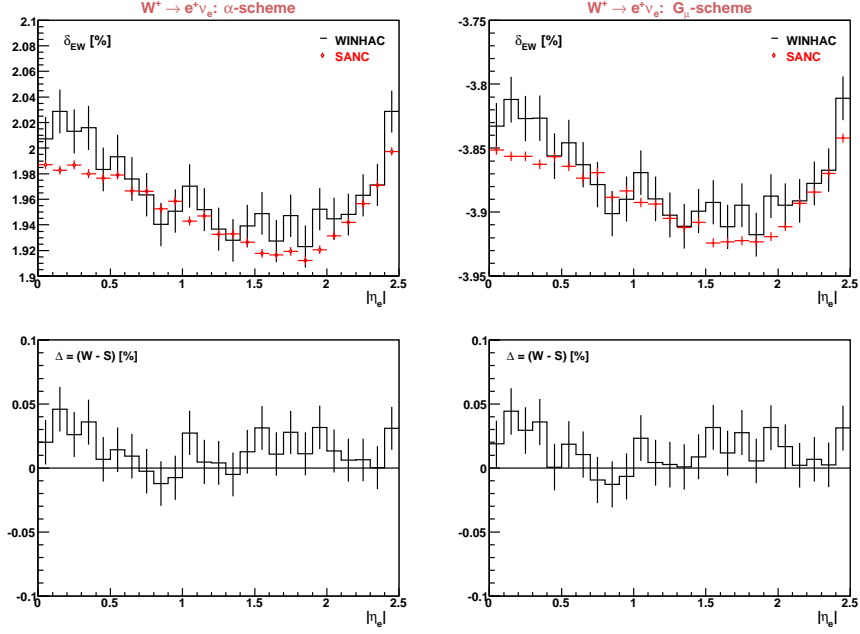}
\end{center}
\vspace*{-5mm}
\caption{The EW NLO distributions of $|\eta_\ell|$ from {\tt SANC} (red diamonds) and {\tt WINHAC} (solid lines) for the electron channel in two schemes and their absolute deviations $\Delta={{\rm W}-{\rm S}}$.
}\label{fig:etael1loop}
\end{figure}

We begin the comparisons of the distributions for the electron channel in two schemes for our three
$W$ observables ($\MTW, \pTl$ and $|\eta_\ell|$, Figs.~\ref{fig:mtel1loop}--\ref{fig:etael1loop}, correspondingly)
with the ``simplified bare'' cuts. The two upper figures show the quantity 
{$\delta_{\rm EW}$} in \%,
while the two lower figures show absolute deviations {$\Delta={\rm W}-{\rm S}$} 
between the two calculations.

As seen, the ${\cal O}(\alpha)$ EW correction {$\delta_{\rm EW}$} is quite large (mainly due to the FSR QED contribution), 
it varies by $18\%$ depending on the scheme. It is shifted
to the larger negative values in the $G_\mu$ scheme and more moderate in the $\alpha(0)$ scheme, the reason for which
the latter was preferred by tuned group of TeV4LHC WS. 
The absolute deviation for both schemes does not exceed $0.1\%$ in the important regions where the cross section is large.
For the $p^e_{\rm T}$ distributions, it varies within $25\%$ but this is an artificial result of applying ``simplified bare'' cuts.
The $\eta_\ell$ distributions are flat and show little biases of the order of a quarter of a per mill. However, most likely VEGAS 
errors are underestimated in the {\tt SANC} results.

\subsubsection{NLO distributions: muon channel} \label{Subsection:NLOmu}
\begin{figure}[!ht]
\begin{center} 
\includegraphics[width=12.0cm]{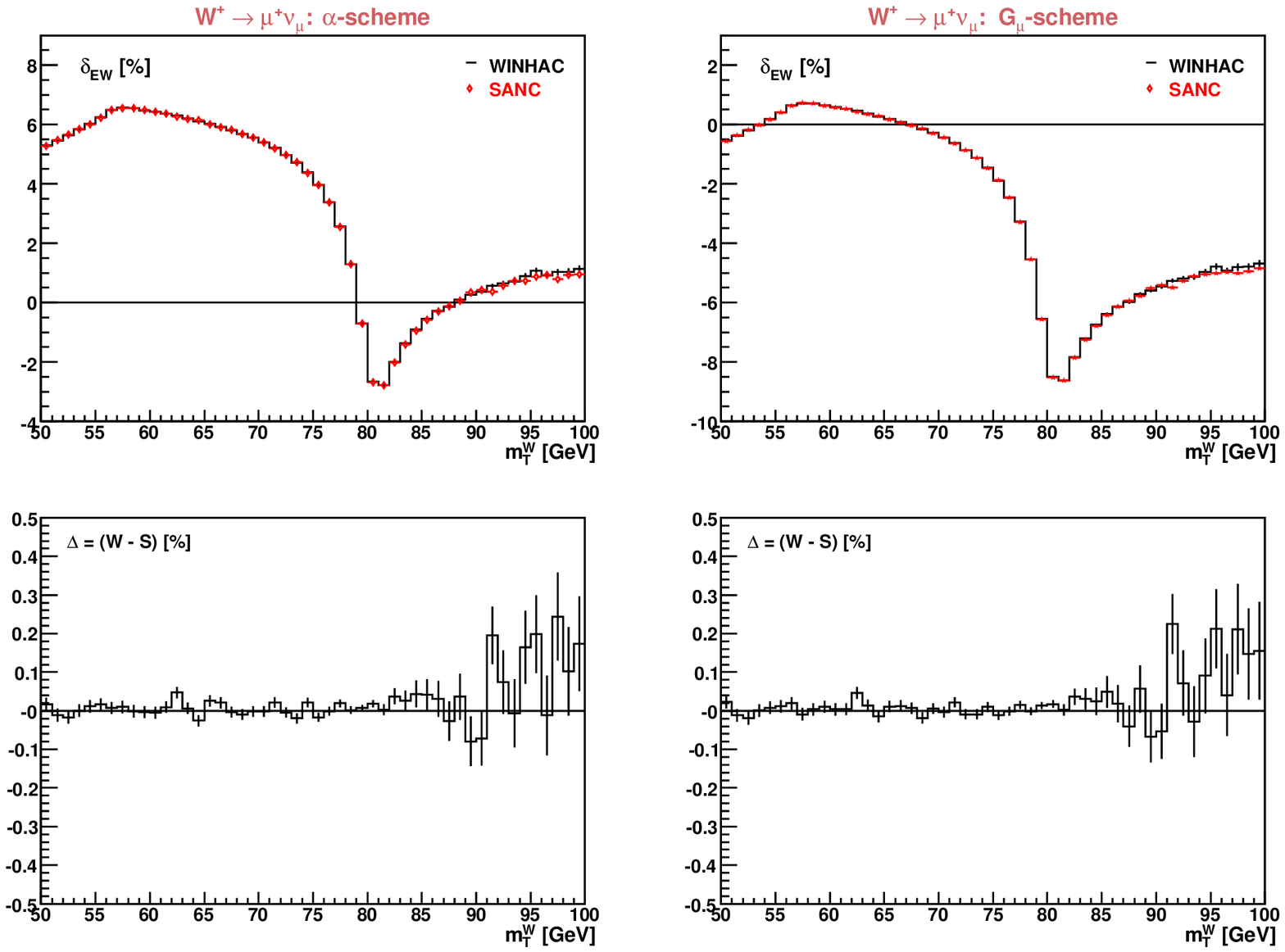}
\end{center}
\vspace*{-5mm}
\caption{The EW NLO distributions of $\MTW$ from {\tt SANC} (red diamonds) and {\tt WINHAC} (solid lines) for the muon channel in two schemes and their absolute deviations $\Delta={{\rm W}-{\rm S}}.$
}\label{fig:mtmu1loop}
\end{figure}

\begin{figure}[!ht]
\begin{center} 
\includegraphics[width=12.0cm]{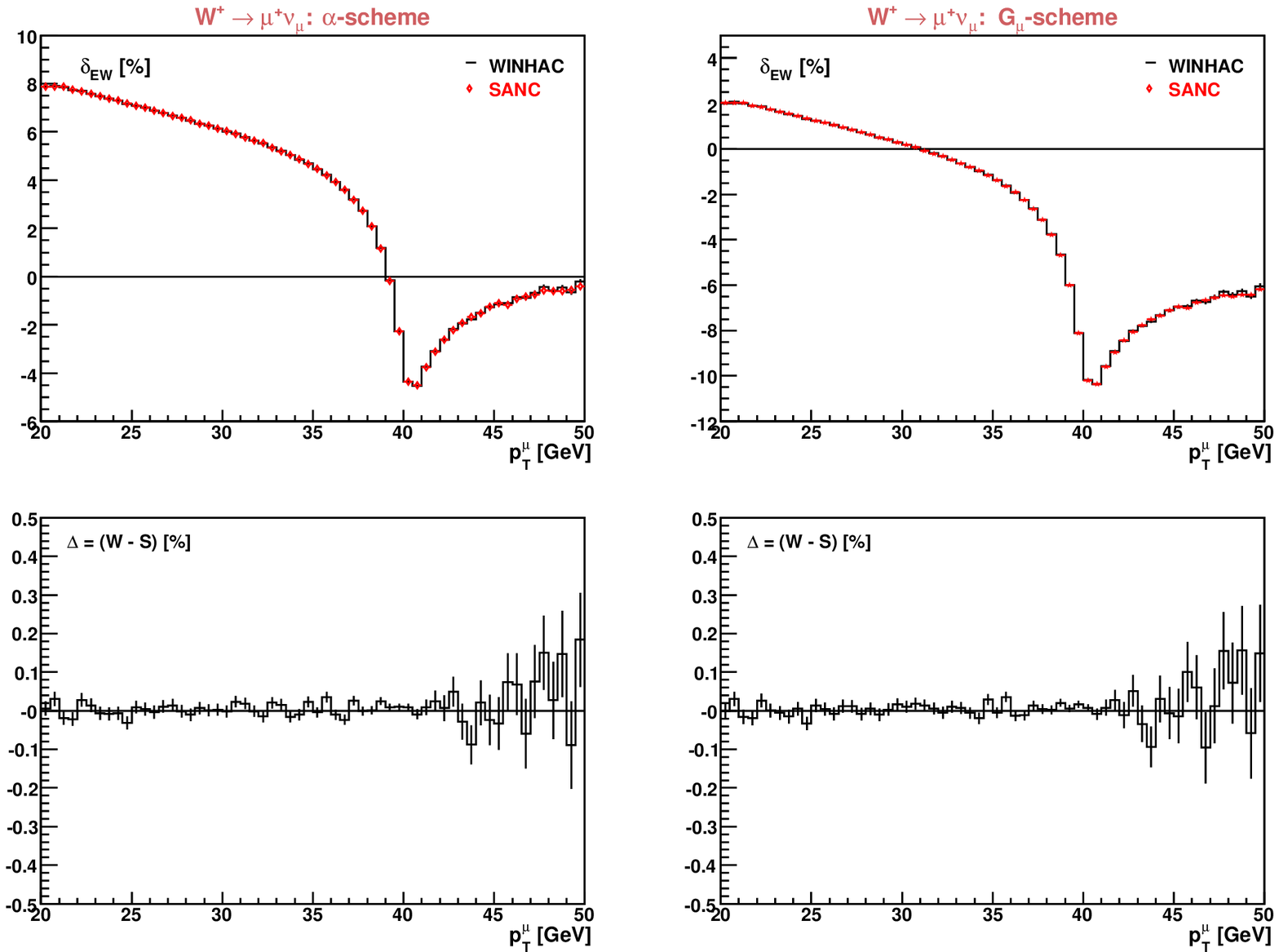}
\end{center}
\vspace*{-5mm}
\caption{The EW NLO distributions of $\pTl$ from {\tt SANC} (red diamonds) and {\tt WINHAC} (solid lines) for the muon channel in two schemes and their absolute deviations $\Delta={{\rm W}-{\rm S}}.$
}\label{fig:ptmu1loop}
\end{figure}

\begin{figure}[!ht]
\begin{center} 
\includegraphics[width=12.0cm]{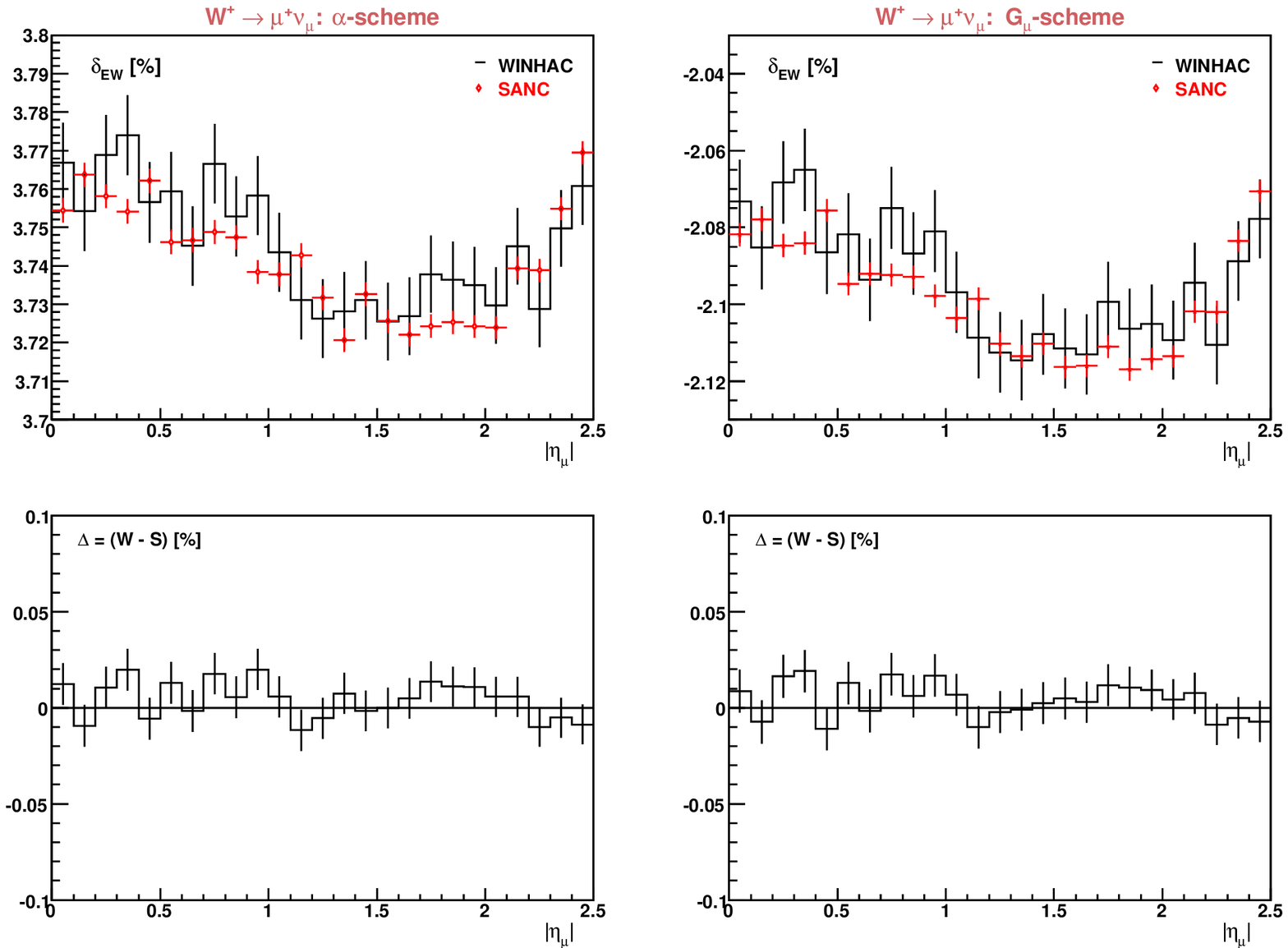}
\end{center}
\vspace*{-5mm}
\caption{The EW NLO distribution of $|\eta_\ell|$ from {\tt SANC} (red diamonds) and {\tt WINHAC} (solid lines) for the muon channel in two schemes and their absolute deviations $\Delta={{\rm W}-{\rm S}}.$
}\label{fig:etamu1loop}
\end{figure}

We continue the comparisons for muon channels in two schemes for the same three
$W$ observables ($\MTW, \pTl$ and $|\eta_\ell|$) with the ``simplified bare'' cuts. The results are presented in 
Figs.~\ref{fig:mtmu1loop}--\ref{fig:etamu1loop}, respectively.
Again, the two upper figures show EW NLO correction {$\delta_{\rm EW}$} in \%,
and the two lower figures show absolute deviations {W$-$S} between the two calculations.
Here the absolute deviations in statistically saturated regions do not exceed $0.05\%$ and in average is of the order of $0.025\%$. 
For the muon channel both calculations are statistical consistent and no evident biases are observed.

It is important to emphasis that biases could be present, in principle, due to finite muon mass,
which treatment in two calculations is not identical: for the muon channel {\tt SANC} uses fully massive formulae 
for all contributions while {\tt WINHAC} uses a mixed approach -- electroweak virtual and soft real-photon corrections are calculated
in the massless fermion approximation, while massive fermions are kept in hard real-photon radiation.

\clearpage

\subsubsection{Weak corrections\label{sec:weak}} \label{subsec:weak}

Here we discuss the ``purely weak'' corrections which are defined as
\begin{eqnarray}
\delta_{\rm weak}&=&\delta^{\rm EW}_{\rm softvirt}
\,-\,\delta^{\rm YFS}_{\rm softvirt}\,,
\label{eq:delta_weak}
\end{eqnarray}
where
\begin{eqnarray}
\delta^{\rm YFS}_{\rm softvirt}\,=\,\delta^{\rm YFS}_{\rm ISR}
\,+\,\delta^{\rm YFS}_{\rm Int}
\,+\,\delta^{\rm YFS}_{\rm FSR}\,,
\label{eq:delta_yfs}
\end{eqnarray}
with three contributions given by Eqs.~(\ref{eq:delta_yfs_subtr},\ref{eq:delta_yfs_int}--\ref{eq:delta_yfs_fsr}).
The contribution $\delta^{\rm EW}_{\rm softvirt}$ includes the 1-loop EW
corrections plus the real soft-photon correction and is provided by the
{\tt SANC} modules.
This definition is free of any regularization scales.

From the Table~\ref{tab:comp_weak} one sees, that
for the electron channel the agreement is very good, while for the
muon channel we observe the systematic differences of about $0.007\%$.
This can be attributed to different treatment of the muon mass in
the two programs: {\tt SANC} uses the fully massive formulae  
while {\tt WINHAC} uses the massless-lepton approximation for these corrections.
The ``weak'' corrections in the $\alpha$-scheme are quite sizable, $\sim 6\%$, because of the light-fermion 
loop contributions, $\sim \ln(\hat{s}/m_f^2)$, to the $W$ self-energy correction. 
Such contributions drop out in the {$G_{\mu}$-scheme} making the ``weak'' corrections 
much smaller, $\sim 0.1\%$.

\begin{table}[!ht]
\begin{center}
\begin{tabular}{||l|c|c||}
\hline
\hline
\multicolumn{3}{||c||}{$\delta_{\rm weak}$ [\%] }
 \\ \hline\hline
\multicolumn{3}{||c||}{\bf LHC, $p p \to W^+\,+\,X \to e^+ \nu_e\,+\,X$}                       \\ \hline
                           & {$\alpha$-scheme}       & {$G_{\mu}$-scheme}        \\ \hline
 {\tt SANC}                & $5.7223(2)$              &$-0.1175(2)$\\
 {\tt WINHAC}              & $5.7220(3)$              &$-0.1177(0)$\\
\hline\hline
\multicolumn{3}{||c||}{\bf LHC, $p p \to W^+\,+\,X \to \mu^+ \nu_\mu\,+\,X$}                   \\ \hline
                          & {$\alpha$-scheme}        & {$G_{\mu}$-scheme} \\ \hline
 {\tt SANC}               & $5.7286(2)$               &$-0.1109(2)$\\
 {\tt WINHAC}             & $5.7220(2)$               &$-0.1177(0)$\\
\hline
\hline
\end{tabular}
\caption{The tuned comparisons of the ``purely weak'' corrections $\delta_{\rm weak}$ from {\tt SANC} and {\tt WINHAC} 
for the simplified bare cuts. The statistical errors of the Monte Carlo integration are given in parentheses.} 
\label{tab:comp_weak}
\end{center}
\end{table}

\begin{figure}[!ht]
\begin{center} 
\includegraphics[width=11.8cm]{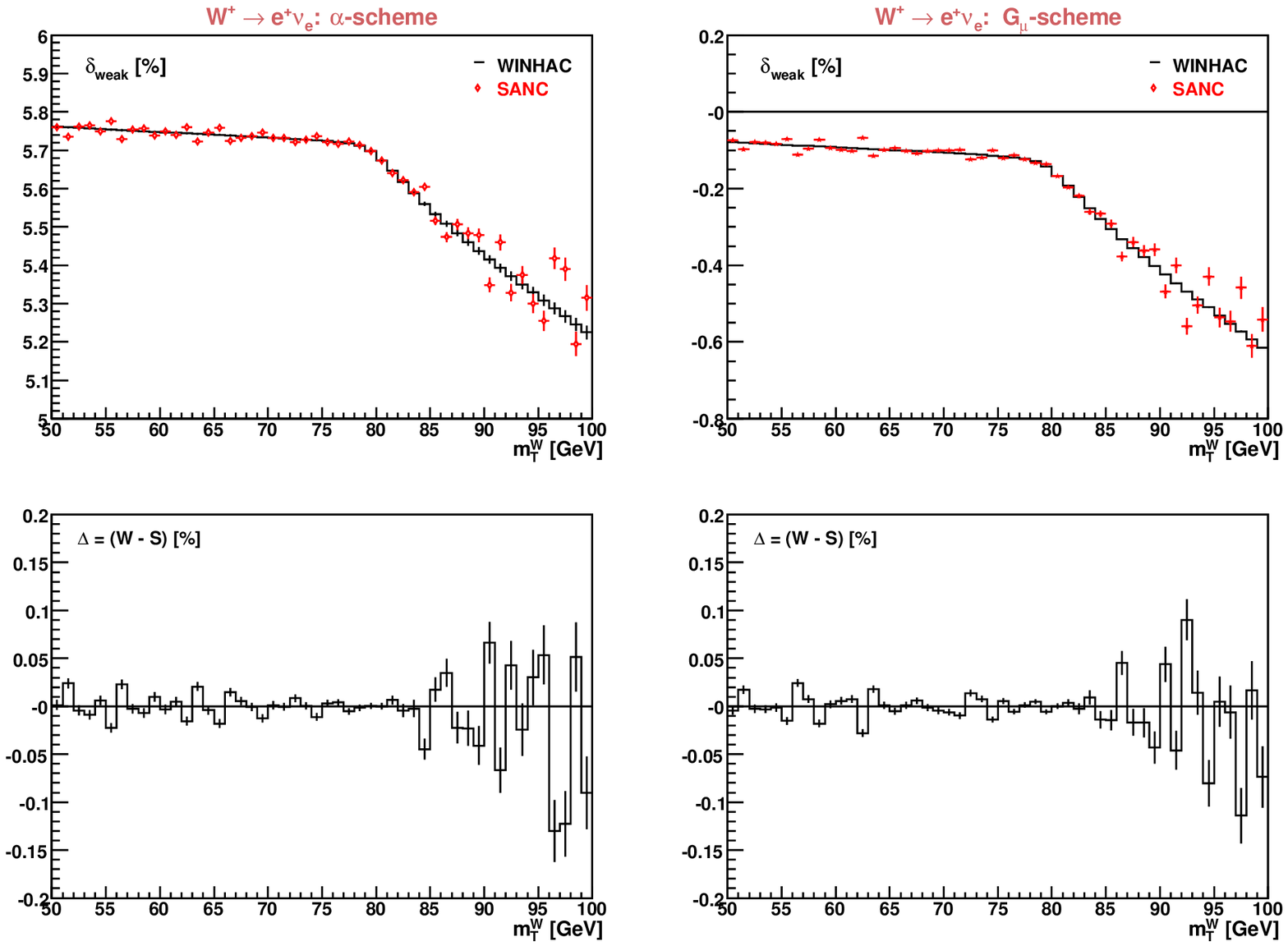}
\end{center}
\vspace*{-5mm}
\caption{The ``weak'' correction distributions of $\MTW$ from {\tt SANC} (red diamonds) and {\tt WINHAC} (solid lines) for the electron channel in two schemes 
and their absolute deviations $\Delta={{\rm W}-{\rm S}}.$
}\label{fig:mtelweak}
\end{figure}

\begin{figure}[!ht]
\begin{center} 
\includegraphics[width=11.8cm]{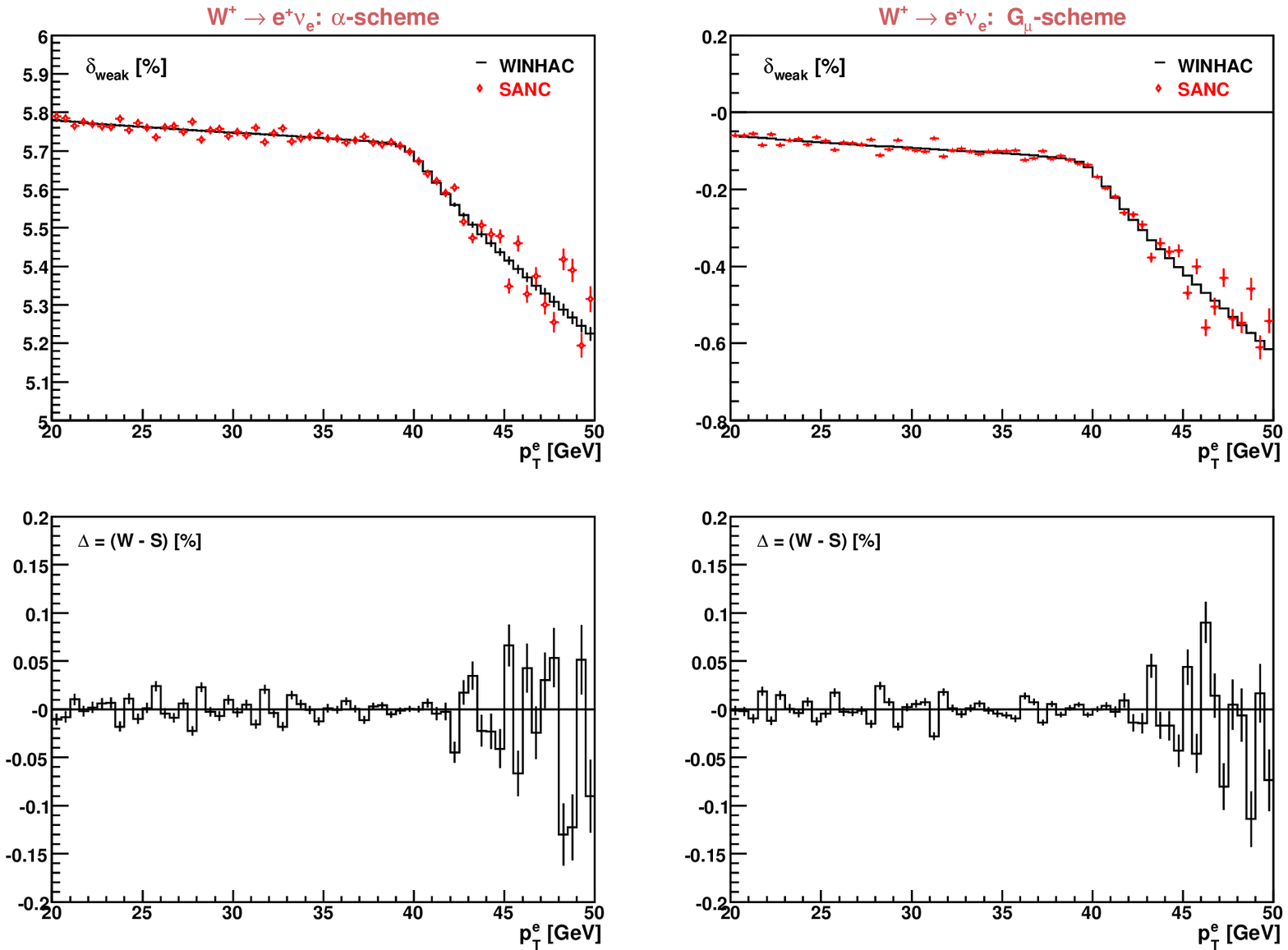}
\end{center}
\vspace*{-5mm}
\caption{The ``weak'' correction distributions of $\pTl$ from {\tt SANC} (red diamonds) and {\tt WINHAC} (solid lines) for the electron channel in two schemes 
and their absolute deviations $\Delta={{\rm W}-{\rm S}}.$
}\label{fig:ptelweak}
\end{figure}

\begin{figure}[!ht]
\begin{center} 
\includegraphics[width=11.8cm]{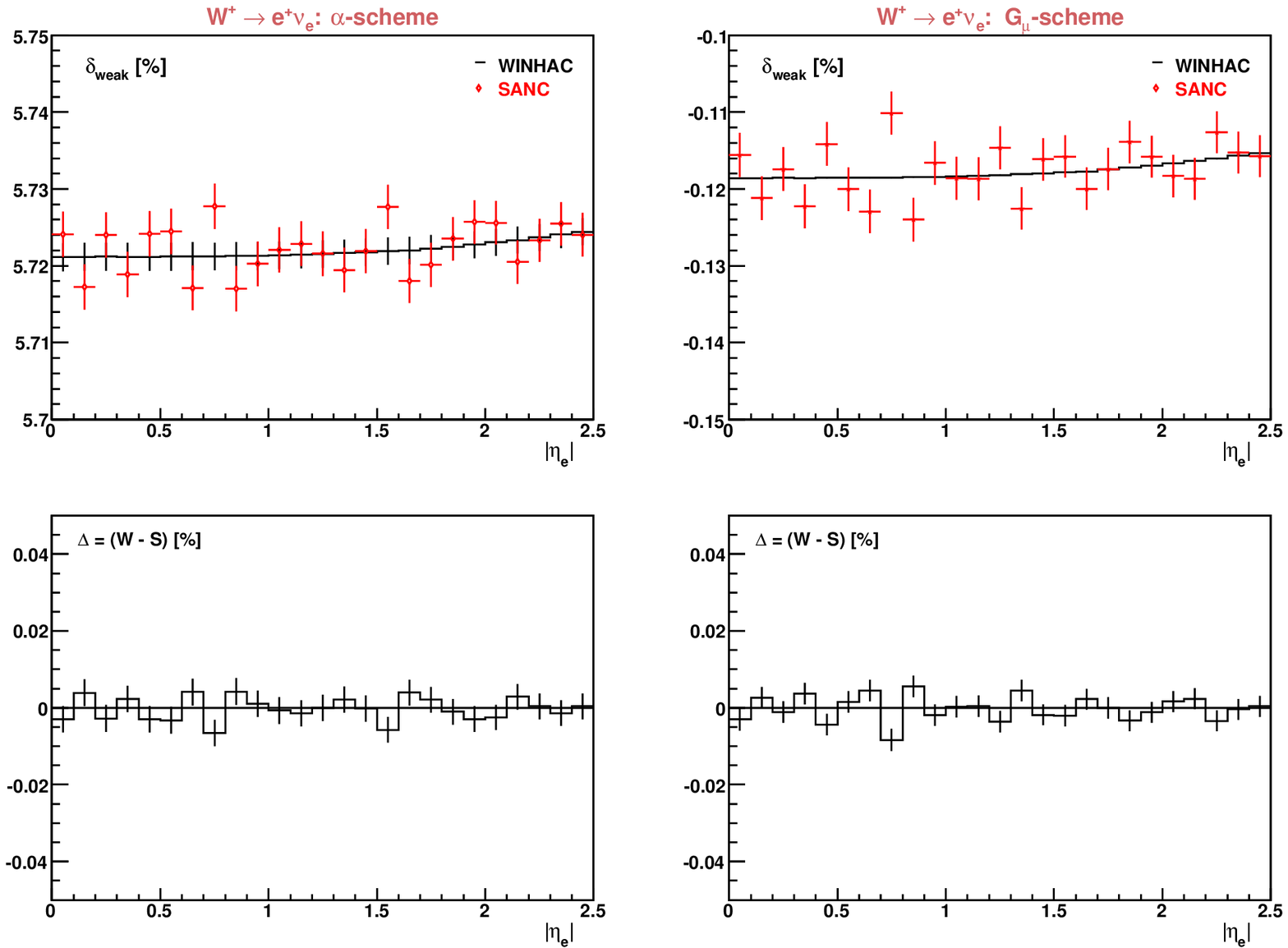}
\end{center}
\vspace*{-5mm}
\caption{The ``weak'' correction distributions of $|\eta_\ell|$ from {\tt SANC} (red diamonds) and {\tt WINHAC} (solid lines) for the electron channel 
in two schemes and their absolute deviations $\Delta={\rm W}-{\rm S}.$
}\label{fig:etaelweak}
\end{figure}

\begin{figure}[!ht]
\begin{center} 
\includegraphics[width=11.8cm]{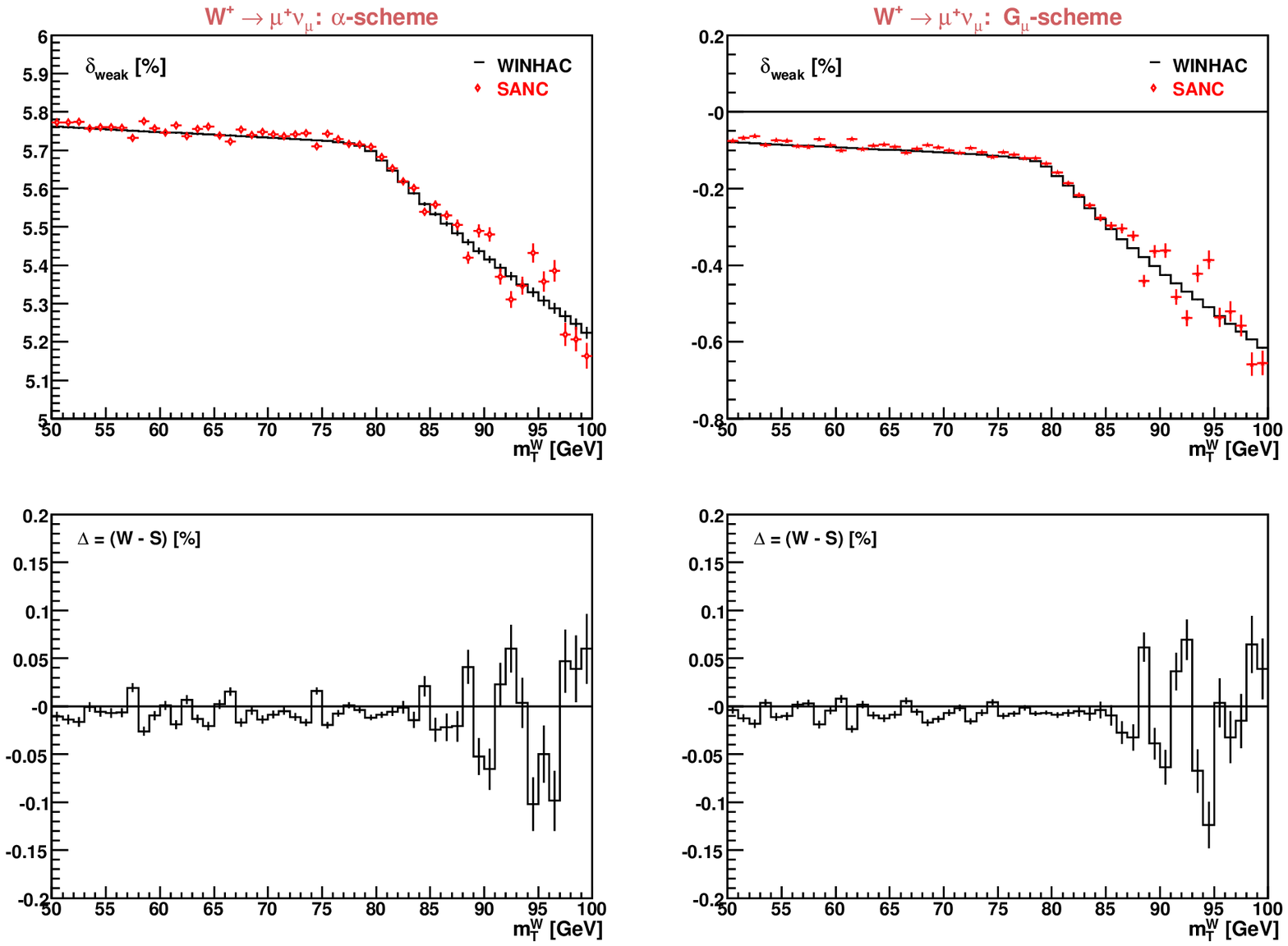}
\end{center}
\vspace*{-5mm}
\caption{The ``weak'' correction distributions of $\MTW$ from {\tt SANC} (red diamonds) and {\tt WINHAC} (solid lines) for the muon channel in two schemes 
and their absolute deviations $\Delta={{\rm W}-{\rm S}}.$
}\label{fig:mtmuweak}
\end{figure}

\begin{figure}[!ht]
\begin{center} 
\includegraphics[width=11.8cm]{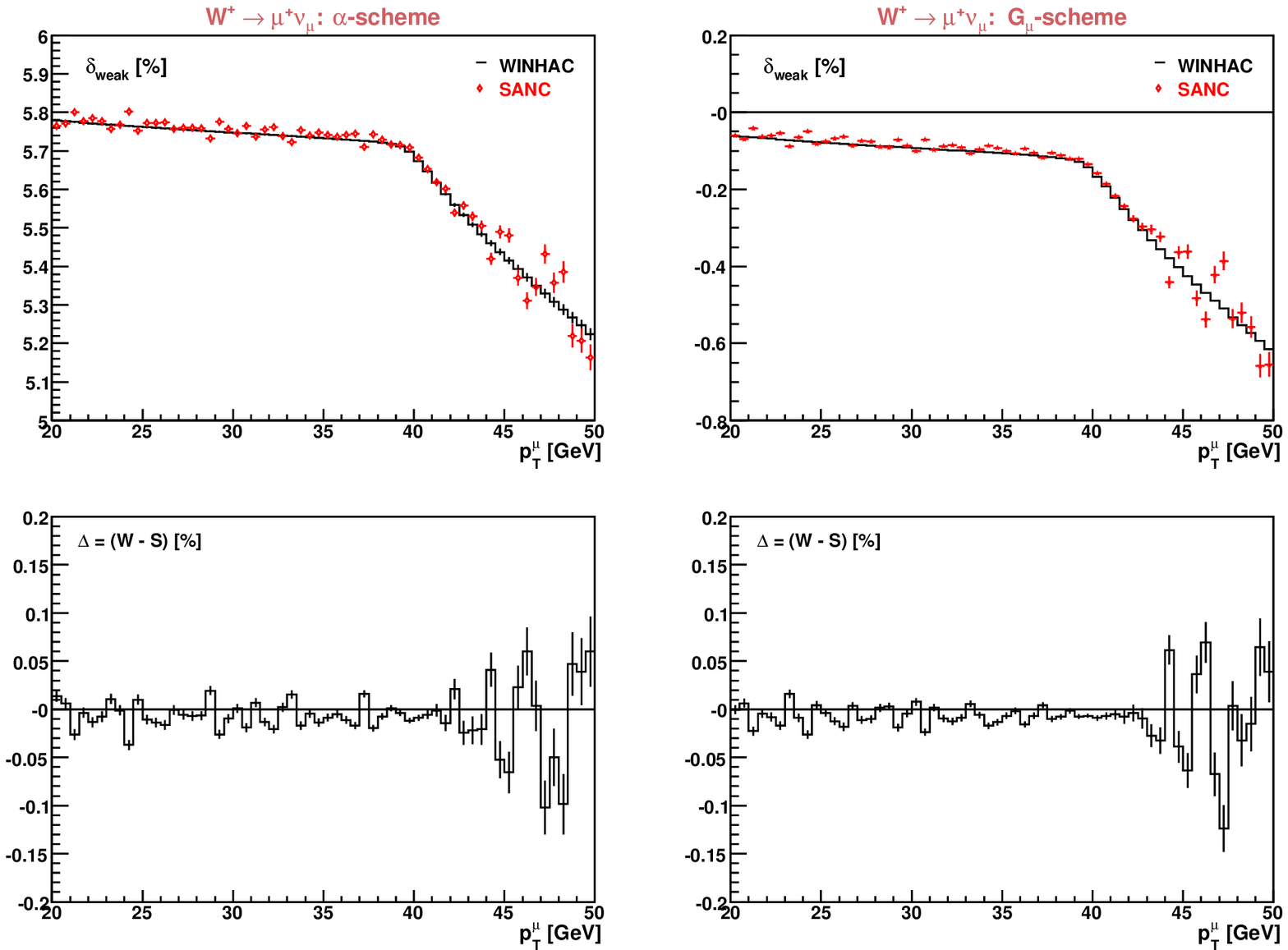}
\end{center}
\vspace*{-5mm}
\caption{The ``weak'' correction distributions of $\pTl$ from {\tt SANC} (red diamonds) and {\tt WINHAC} (solid lines) for the muon channel in two schemes 
and their absolute deviations $\Delta={{\rm W}-{\rm S}}.$
}\label{fig:ptmuweak}
\end{figure}

\begin{figure}[!ht]
\begin{center} 
\includegraphics[width=11.8cm]{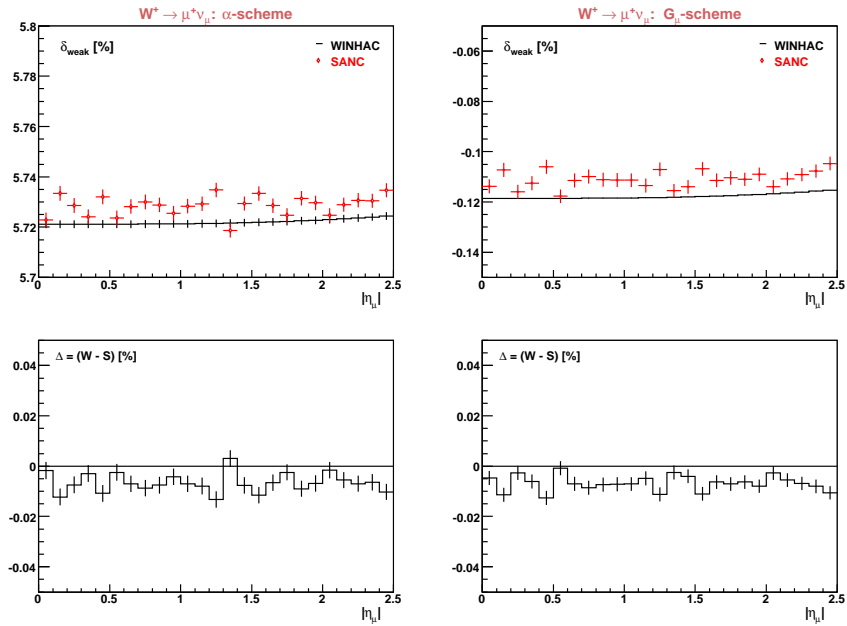}
\end{center}
\vspace*{-5mm}
\caption{The ``weak'' correction distributions of $|\eta_\ell|$ from {\tt SANC} (red diamonds) and {\tt WINHAC} (solid lines) for the muon channel 
in two schemes and the absolute deviations $\Delta={\rm W}-{\rm S}.$
}\label{fig:etamuweak}
\end{figure}

In Figs.~\ref{fig:mtelweak}--\ref{fig:etamuweak} we show the distributions of the ``weak'' corrections
and absolute deviations between the two calculations.
The figures show agreement at the level $0.01\%$. In some cases the biases of the same order are seen.
Again, this might be a consequence of underestimation of errors by VEGAS.
In the muon channel, the observed deviations at the level of $0.01\%$ can
be attributed again to different treatment of the muon mass in the two
programs. 

\clearpage
\section{Conclusions\label{sec:concl}}

The main priority of the development of {\sanc} as a HEP tool for the LHC is to create the SSFM
for the EW corrections at one-loop level to be used in existing MC event generators.

The goals of this work were:
(a) to integrate CC DY SSFM into the Monte Carlo event 
generator {\tt WINHAC} and 
(b) to check thoroughly the stability of numbers for simple distributions
by comparisons of the {\tt WINHAC} generated results with those provided by
 the recently created {\sanc} CC DY integrator. 
In this paper we have concentrated on presenting the numerical tests of
the implementation of the above EW corrections in {\tt WINHAC}, while the
details on this implementation will be given elsewhere.  

The main and very important conclusion of this paper is that we have reached 
the agreement between the {\tt WINHAC} MC event generator 
and the {\sanc} MC integrator 
for the ${\cal O}(\alpha)$ EW corrections to the charged-current 
Drell--Yan process at the sub-per-mill level, both for the inclusive cross
section and for the main distributions.
Thus, our above goals have been achieved.

Another important conclusion is that the MC event generator
{\tt WINHAC} can now be used for precision simulations of the charged-current 
Drell--Yan process at the LHC including the ${\cal O}(\alpha)$ EW corrections.
It can also serve as a benchmark for testing other MC programs for this process.

The next step on this road would be a similar implementation of the {\sanc} modules in the neutral-current Drell--Yan MC event generator {\tt ZINHAC},
being under development now.

\vspace*{10mm}

\leftline{\Large\bf Acknowledgements}

\vspace*{2mm}

The authors are grateful to A.~Arbuzov, P.~Christova and R.~Sadykov for numerous discussions of related issues.
Two of us (D.B. and L.K.) are cordially indebted to S.~Jadach and Z.~Was for offering
us an opportunity of encouraging common work with scientists of IFJ Krakow in May--June 2008 and to the IFJ
directorate for hospitality which was extended to us in this period, when the major part of this study was done. 
We also acknowledge the hospitality of the CERN PH Theory Unit 
where this work was finalized.


\providecommand{\href}[2]{#2}
\end{document}